\documentclass[conference]{IEEEtran}
\usepackage[T1]{fontenc}
\usepackage[latin9]{inputenc}
\usepackage{amsthm}
\usepackage{amsmath}
\usepackage{amssymb}
\usepackage{color}

\allowdisplaybreaks
\newtheorem{theorem}{Theorem}

\newtheorem{lemma}[theorem]{Lemma}

\def\MGT{M^{\textrm{GT}}}
\newcommand{\red}[1]{#1}

\begin{document}

\title{Minimax Risk for Missing Mass Estimation}

\author{
\IEEEauthorblockN{Nikhilesh Rajaraman, Andrew Thangaraj}
\IEEEauthorblockA{Department of Electrical Engineering\\
Indian Institute of Technology Madras\\
Chennai 600036, India\\
andrew@ee.iitm.ac.in
}
\and
\IEEEauthorblockN{Ananda Theertha Suresh}
\IEEEauthorblockA{Google Research\\
New York, USA\\
theertha@google.com}
}

\maketitle

\begin{abstract}
The problem of estimating the missing mass or total probability of
unseen elements in a sequence of $n$ random samples is
considered under the squared error loss function. The worst-case risk
of the popular Good-Turing estimator is shown to be between $0.6080/n$
and $0.6179/n$. The minimax risk is shown to be lower bounded by
$0.25/n$. This appears to be the first such published result on minimax
risk for estimation of missing mass, which has several practical and
theoretical applications.
\end{abstract}

\section{Introduction}

Given independent samples from an unknown distribution, missing mass
estimation asks for the sum of the probability of the unseen
elements. Missing mass estimation is a basic problem in statistics and
has wide applications in several fields ranging from language
modeling~\cite{GaleS95,ChenG96} to ecology~\cite{ChaoL92}.
Perhaps the most used missing mass estimator is the Good-Turing
estimator which was proposed in a seminal paper by I. J. Good and Alan
Turing in 1953~\cite{doi:10.1093/biomet/40.3-4.237}. The Good-Turing
estimator is used in support
estimators~\cite{ChaoL92}, entropy estimators~\cite{VuYK07} and unseen species
estimators~\cite{ShenCL03}. To describe the estimator and the results,
we need a modicum of nomenclature.

Let $p$ be an underlying unknown distribution over an unknown domain
$\mathcal{X}$. Let $X^{n}\triangleq(X_{1},X_{2},\ldots,X_{n})$ be $n$
independent samples from $p$. For $x \in \mathcal{X}$, let $N_x(X^n)$
be the number of appearances of $x$ in $X^n$. Upon observing $X^n$,
our goal is to estimate the missing mass
\begin{equation}
M_{0}(X^{n})\triangleq\sum_{u\in\mathcal{X}}p(u)\mathbb{I}(N_{u}(X^{n})=0),\label{eq:1}
\end{equation}
where $\mathbb{I}(\cdot)$ denotes the indicator function. For example, if $\mathcal{X} = \{a,b,c,d\}$
and $X^3 = b \, c \, b$, then $M_0(X^3) = p(a) + p(d)$.  The above
sampling model for estimation is termed the multinomial model.  We
note that $1-M_0(X^n)$ is often referred as sample coverage in the
literature~\cite{CCG12}.

An estimator for missing mass $\hat{M}_0(X^n)$ is a mapping from
$\mathcal{X}^n \to [0,1]$.  For a distribution $p$, the $\ell^2_2$ risk of the estimator $\hat{M}_0(X^n)$ is
\[
R_n(\hat{M}_0, p) \triangleq E_{X^n \sim p} [(\hat{M}_0(X^n) - M_0(X^n))^2],
\]
and the worst-case risk over all distributions is 
\[
R_n(\hat{M}_0) \triangleq \max_p R_n(\hat{M}_0, p),
\]
and minimax mean squared loss or minimax risk is
\[
R^*_n = \min_{\hat{M}_0} R_n(\hat{M}_0).
\]
The goal of this paper is to characterize $R^*_n$. 

\subsection{Good-Turing estimator and previous results}
Let
\[
\Phi_{i}(X^{n})\triangleq\sum_{u\in\mathcal{X}}\mathbb{I}(N_{u}(X^{n})=i)
\]
denote the number of symbols that have appeared $i$ times in $X^{n}$,
$1\le i\le n$. For example, if $X^3 = a, b, c$, then $\Phi_1 = 3$ and 
$\Phi_i = 0$ for all $i > 1$. The Good-Turing estimator~\cite{doi:10.1093/biomet/40.3-4.237} for the missing mass is
\[
M^{\textrm{GT}}(X^n) \triangleq \frac{\Phi_1(X^n)}{n}.
\]
One of the first theoretical analysis of the Good-Turing estimator was
in \cite{McAllester:2000:CRG:648299.755182}, where it was shown that
\begin{equation}
\left|E\left[M^{\textrm{GT}}(X^{n})-M_{0}(X^{n})\right]\right|\le\frac{1}{n}.\label{eq:3}
\end{equation}
This shows that the bias of the Good-Turing estimator falls as $1/n$. They further showed that 
with probability $\geq 1- \delta$,
\[
\left \lvert M^{\textrm{GT}}(X^{n})-M_{0}(X^{n}) \right \rvert 
\leq \frac{2}{n} + \sqrt{\frac{2\ln (3/\delta)}{n}} \left( 1+ 2 \ln (3n/\delta) \right).
\]
Various properties of the Good-Turing estimator and several variations of
it have been analyzed for distribution estimation and
compression~\cite{OrlitskySZ03, DrukhM04, WagnerVK06,
  wagner2007better, OhannessianD12, AcharyaJOS13, OS15}. Several
concentration results on missing mass estimation are also
known~\cite{berend2013concentration, ben2017concentration}.  Despite
all this work, the risk of the Good-Turing estimator and the
minimax risk of missing mass estimation have still not been
conclusively established.

\subsection{New results}
 Unlike parameters of a distribution, missing mass itself is a
 function of the observed sample and that makes finding the exact minimax
 risk difficult. 

We first analyze the risk of the Good-Turing
estimator and show that for any distribution $p$,
\begin{align*}
R_n(M^{\textrm{GT}}, p )&=
 \frac{1}{n}E\left[\frac{2\Phi_2}{n}+\frac{\Phi_1}{n} \left(1- \frac{\Phi_1}{n} \right)
  \right] + o \left( \frac{1}{n}\right),
\end{align*}
where $\Phi_i$ is abbreviated notation for $\Phi_i(X^n)$. By
maximizing the RHS in the first equation above over all distributions, in 
Theorem~\ref{thm:4}, we show that 
\[
\frac{0.6080}{n}+ o\left(\frac{1}{n} \right) \leq R_n(M^{\textrm{GT}}) \leq \frac{0.6179}{n} + o\left( \frac{1}{n}\right).
\]
We note that under the multinomial model, the numbers of occurrences
of symbols are correlated, and this makes finding the worst case
distribution for the Good-Turing estimator difficult. 

 We then prove estimator-independent information-theoretic lower
 bounds on $R^*_n$ using two approaches. We first compute the lower
 bound via Dirichlet prior approach~\cite{Krichevskiy98}. In
 Lemma~\ref{lem:Prior3}, we show that
\[
R^*_n \geq \frac{4}{27n}.
\]
We then improve the constant by reducing the problem of missing mass
estimation to that of distribution estimation. In particular, in
Theorem~\ref{thm:lower_dist}, we show that
\[
R^*_n \geq \frac{1}{4n} + o\left( \frac{1}{n}\right)
\]
Combining the lower and upper bounds, we get
\[
\frac{0.25}{n} + o\left( \frac{1}{n}\right) \leq R^*_n  \leq \frac{0.6179}{n} + o\left( \frac{1}{n}\right),
\]
Finding the exact minimax risk for the missing mass estimation problem
remains
an open question.

The rest of the paper is organized as follows. In
Section~\ref{sec:upper}, we analyze the Good-Turing estimator. In
Section~\ref{sec:prior}, we use Dirichlet prior approach to obtain
lower bounds and in Section~\ref{sec:dist} we obtain lower bounds via
reduction.

\section{Risk of Good-Turing Estimator}
\label{sec:upper}

The analysis of \cite{McAllester:2000:CRG:648299.755182} can be extended
to characterize the risk of the Good-Turing estimator
for missing mass. The squared error of the Good-Turing estimator $\MGT(X^n)$ can be written down as follows:
\begin{align}
&\left(\MGT(X^{n})-M_{0}(X^{n})\right)^{2}\nonumber \\
&\quad =\left(\sum_{u\in\mathcal{X}}\frac{1}{n}\mathbb{I}(N_{u}=1)-p(u)\mathbb{I}(N_{u}=0)\right)\nonumber\\
&\qquad\qquad\left(\sum_{v\in\mathcal{X}}\frac{1}{n}\mathbb{I}(N_{v}=1)-p(v)\mathbb{I}(N_{v}=0)\right)\nonumber \\
&\quad=\frac{1}{n^2}\sum_{u,v\in\mathcal{X}}\bigg(\mathbb{I}(N_{u}=1)\mathbb{I}(N_{v}=1)\nonumber\\
&\qquad\qquad\qquad-2np(u)\mathbb{I}(N_{u}=0)\mathbb{I}(N_{v}=1)\nonumber\\
&\qquad\qquad\qquad+n^2p(u)p(v)\mathbb{I}(N_{u}=0)\mathbb{I}(N_{v}=0)\bigg)\label{eq:5}
\end{align}
For $u,v\in\mathcal{X}$, $E[\mathbb{I}(N_{u}=i)I(N_{v}=j)]=\mathbb{P}(N_{u}=i,N_{v}=j)$.
Using the notation $P_{n}(i,j)=\mathbb{P}(N_{u}(X^{n})=i,N_{v}(X^{n})=j)$,
we get 
\begin{align}
R_n(\MGT,p) &                    =\frac{1}{n^2}\sum_{u,v\in\mathcal{X}}\bigg(P_{n}(1,1)-2np(u)P_{n}(0,1)\nonumber\\
&\qquad\qquad\qquad+n^2p(u)p(v)P_{n}(0,0)\bigg).\label{eq:6}
\end{align}
The probability $P_{n}(i,j)$ can be written down as 
\begin{equation}
P_{n}(i,j)=\begin{cases}
\binom{n}{i\ j}\ p(u)^{i}p(v)^{j}(1-p(u)-p(v))^{n-i-j},\; u\ne v,\\[10pt]
\binom{n}{i}\ p(u)^{i}(1-p(u)^{n-i},\; u=v,i=j,
\end{cases}\label{eq:7}
\end{equation}
where $\binom{n}{i\ j}=\frac{n!}{i!j!(n-i-j)!}$ and
$\binom{n}{i}=\frac{n!}{i!(n-i)!}$. The summation in (\ref{eq:6}) is first split into two cases:
$u\ne v$ and $u=v$. Denoting $P(u,v)=p(u)p(v)(1-p(u)-p(v))^{n-2}$,
we have, for $u\ne v$,
\begin{align*}
  p(u)p(v)P_n(0,0)&=(1-p(u)-p(v))^2P(u,v),\\
  p(u)P_n(0,1)&=n(1-p(u)-p(v))P(u,v),\\
  P_n(1,1)&=n(n-1)P(u,v).
\end{align*}
For $u=v$, observe that
$P_n(0,1)=0$. Using the above observations, the summation in \eqref{eq:6} simplifies to
\begin{align}
&R_n(\MGT,p)=\frac{1}{n}\sum_{\substack{u,v\in\mathcal{X}\\v\ne u}}P(u,v)\bigg[n\big(p(u)+p(v)\big)^2-1\bigg]\nonumber \\
&\ +\frac{1}{n}\sum_{u\in\mathcal{X}}
   \bigg[p(u)(1-p(u))^{n-1}+np(u)^{2}(1-p(u))^{n}\bigg].\label{eq:9}
\end{align}
The following lemma is useful in bounding certain terms in the first summation above as a function
of $n$, independent of the unknowns $\mathcal{X}$ and $p$. 
\begin{lemma}
\label{lem:2}For $i\ge1$, $j\ge1$, 
\[
\sum_{u,v\in\mathcal{X}, \red{u \neq v}}p(u)^{i}p(v)^{j}(1-p(u)-p(v))^{n}\le\dfrac{(i-1)!\ (j-1)!\ n!}{(n+i+j-2)!}.
\]
\end{lemma}
\begin{IEEEproof}
Let $X$ and $Y$ be a pair of
independent and identical random variables with marginal distribution
$p$. Define a random variable $T(X,Y)$, whose value $T(u,v)=0$ for
$u=v$ and, for $u\ne v$, 
$$T(u,v)=\binom{n+i+j-2}{i-1\ j-1}p(u)^{i-1}p(v)^{j-1}(1-p(u)-p(v))^{n}.$$
We see that $T(X,Y)$ is a probability for $X\ne Y$, and that it takes values
in $[0,1]$ in all cases. Therefore, its expectation
\begin{align*}
& \red{ E[T(X,Y)] = \sum_{u, v \in \mathcal{X} \atop u \ne v}  p(u)p(v) T(u,v)} \\
& \red{=}\sum_{u,v\in\mathcal{X}\atop u\ne
  v}\binom{n+i+j-2}{i-1\ j-1}p(u)^ip(v)^j(1-p(u)-p(v))^n\le 1,
\end{align*}
which concludes the proof.
\end{IEEEproof}
A useful univariate version of Lemma \ref{lem:2} is the following.
\begin{lemma}
\label{lem:1} For $i\ge1$, 
\[
\sum_{u\in\mathcal{X}}p(u)^{i}(1-p(u))^{n}\le\dfrac{(i-1)!\ n!}{(n+i-1)!}.
\]
\end{lemma}
\begin{IEEEproof}
  For $X\sim p$, define $T(X)=\binom{n+i-1}{i-1}p(X)^{i-1}(1-p(X))^n$
  and follow the proof of Lemma \ref{lem:2}.
\end{IEEEproof}
Using Lemma \ref{lem:2}, observe that
\begin{align}
  \label{eq:20}
  \sum_{u,v\in\mathcal{X}\red{,u \neq v}}P(u,v)(p(u)+p(v))^2=o(1/n).
\end{align}
Therefore, the risk can be written as
\begin{align}
R_n(\MGT,p)&=\frac{1}{n}\bigg[\sum_{u\in\mathcal{X}}
   p(u)(1-p(u))^{n-1}-\sum_{\substack{u,v\in\mathcal{X}\\v\ne u}}P(u,v)\nonumber\\
&\ +\sum_{u\in\mathcal{X}} np(u)^{2}(1-p(u))^{n}\bigg]+o(1/n).  \label{eq:21}
\end{align}
The summation terms above can be rewritten as follows: 
\begin{align}
\sum_{u\in\mathcal{X}}p(u)(1-p(u))^{n-1}&=E\bigg[\frac{\Phi_{1}(X^{n})}{n}\bigg].\label{eq:23}\\
\sum_{u\in\mathcal{X}}np(u)^2(1-p(u))^{n}&=\frac{2}{n-1}\sum_{u\in\mathcal{X}}P_n(2,0)(1-p(u))^2\nonumber\\
&\overset{(a)}{=}\frac{2}{n-1}\sum_{u\in\mathcal{X}}P_n(2,0)\pm
  {o\left(\frac{1}{n}\right)}\nonumber\\
&=E\left[\frac{2\Phi_2(X^n)}{n}\right]\pm {o\left(\frac{1}{n}\right)},\label{eq:24}
\end{align}
where $(a)$ follows using Lemma \ref{lem:1}.
\begin{align}
&\sum_{\substack{u,v\in\mathcal{X}\\v\ne
  u}}P(u,v)=\frac{1}{n(n-1)}\sum_{\substack{u,v\in\mathcal{X}\\v\ne
  u}}P_n(1,1)\nonumber\\
&=\frac{1}{n(n-1)}E\bigg[\sum_{\substack{u,v\in\mathcal{X}\\v\ne
  u}}\mathbb{I}(N_{u}(X^{n})=1)\mathbb{I}(N_{v}(X^{n})=1)\bigg]\nonumber\\
&=E\bigg[\frac{1}{n(n-1)}\Phi_1(X^n)(\Phi_1(X^n)-1)\bigg]\nonumber\\
&=E\left[\frac{\Phi^2_{1}(X^n)}{n}\right]\pm o(1).\label{eq:25}
\end{align}
Using the above expressions in \eqref{eq:21}, we get the following characterization of the
risk.
\begin{theorem}
  The risk of the Good-Turing estimator under squared error loss
  satisfies 
  \begin{equation}
    \label{eq:17}
    R_n(M^{\textrm{GT}}, p)= \frac{1}{n}E\left[\frac{2\Phi_2}{n}+\frac{\Phi_1}{n} \left(1- \frac{\Phi_1}{n} \right)\right]+ {o\left(\frac{1}{n}\right)}.
  \end{equation}
\label{thm:riskGT}
\end{theorem}
\subsection{Upper bound on risk}
To obtain a tight upper bound on the risk, we start with
the following upper bound on one of the terms in \eqref{eq:21}:
\begin{align}
\sum_{u\in\mathcal{X}}np(u)^{2}(1-p(u))^{n}&\le\sum_{u\in\mathcal{X}}p(u)\left(np(u)e^{-np(u)}\right)\nonumber\\
&\le e^{-1},\label{eq:12}
\end{align}
where the first step follows because $1-x\le e^{-x}$ for a fraction
$x$, and the second step follows because $te^{-t}\le e^{-1}$ for
$t\ge0$. Using \eqref{eq:23}, \eqref{eq:24} and \eqref{eq:12} in
\eqref{eq:21}, an upper bound for the risk of the Good-Turing
estimator is 
\begin{align}
R_n(\MGT,p)&\le\frac{1}{n}E\left[\frac{\Phi_1}{n} \left(1-
             \frac{\Phi_1}{n} \right)\right]+\frac{e^{-1}}{n}\pm
             o\left(\frac{1}{n}\right)\nonumber\\
&\le\frac{0.25+e^{-1}}{n}\pm {o\left(\frac{1}{n}\right)},\label{eq:13}
\end{align}
where the last step follows because $x(1-x)\le 0.25$ for a fraction
$x$. The above constant $e^{-1}+0.25\approx0.6179$ is not best possible,
and could be marginally improved by more careful analysis. However, we
show that the improvement is not significant through a lower bound on
$R_n(\MGT)=\max_pR_n(\MGT,p)$ by picking $p$ to be a suitable uniform distribution.

\subsection{Lower bound on the Good-Turing worst-case risk}
A lower bound can be obtained for the worst case risk of the Good-Turing
estimator by evaluating the risk for the uniform distribution $p_U$ on $\mathcal{X}$. Let $\left|\mathcal{X}\right|=cn$
and $p_U\left(x\right)=\frac{1}{cn}$ for all $x\in\mathcal{X}$, where
$c$ is a positive constant. Using (\ref{eq:21}), we get
\begin{align}
&R_n(\MGT,p_U)= \frac{1}{n}\bigg[\frac{cn\cdot
  n}{(cn)^{2}}\left(1-\frac{1}{cn}\right)^{n} +\frac{cn}{cn}\cdot\left(1-\frac{1}{cn}\right)^{n-1}\nonumber\\
&\quad-\left(\frac{cn}{cn}\cdot\left(1-\frac{1}{cn}\right)^{n-1}\right)^{2}\bigg]+o\left(\frac{1}{n}\right)\nonumber \\
 & \overset{(a)}{=}  \frac{1}{n}\left(\left(\frac{1}{c}+1\right)\left(1-\frac{1}{cn}\right)^{n}-\left(1-\frac{1}{cn}\right)^{2n}\right)+o\left(\frac{1}{n}\right)\nonumber \\
 & \overset{(b)}{=}  \frac{1}{n}\left(\left(\frac{1}{c}+1\right)e^{-\frac{1}{c}}-e^{-\frac{2}{c}}\right)+o\left(\frac{1}{n}\right)\label{eq:14}
\end{align}
where the reasoning for the steps is as follows: 
\begin{enumerate}
\item replacing $\left(1-\frac{1}{cn}\right)^{n-1}$ with $\left(1-\frac{1}{cn}\right)^{n}\left(1+o(1)\right)$.
\item using the fact that $\left(1-\frac{1}{cn}\right)^{n}=e^{-1/c}\left(1+o(1)\right)$.
\end{enumerate}
The coefficient of $\frac{1}{n}$ in (\ref{eq:14}) can be maximized
numerically to obtain a maximum value of $0.6080$ at $c\approx1.1729$.
Hence, from (\ref{eq:13}) and (\ref{eq:14}), we have:
\begin{theorem}
\label{thm:4}
The worst-case risk of the Good-Turing estimator satisfies the
following bounds:
\begin{multline}
\frac{0.6080}{n}+o\left(\frac{1}{n}\right)\leq R_n(\MGT)
 \leq\frac{0.6179}{n}+o\left(\frac{1}{n}\right).
\end{multline}
\end{theorem}
Therefore, the constant in (\ref{eq:13}) is fairly tight.

\section{Lower Bounds on the Minimax Risk}
In this section, we consider lower bounds on the squared error risk of an arbitrary
estimator of missing mass. The main result is that the minimax risk is
lower-bounded by $c/n$ for a constant $c$. Two methods are described
for finding lower bounds - the first one is a Dirichlet prior approach, and the second one
is reduction of the missing
mass problem to a distribution estimation problem.

Both approaches provide the same order of $1/n$ for the lower bound,
but the second reduction approach provides a better
constant. However, the Dirichlet prior approach has significant
potential for further optimization for
better constants, and is an interesting extension of the standard prior
method to the case of estimation of random variables such as missing mass, which depend on both the
distribution $p$ and the sample $X^n$.

\subsection{Lower Bounds via Prior Distributions}
\label{sec:prior}
The first approach is to bound the minimax risk by the average risk obtained
by averaging over a family of distributions with a prior. Let $P$
be a random variable over a family of distributions $\mathcal{P}$,
having an alphabet $\mathcal{X}=\left\{ 0,1,2,\ldots k-1\right\} $.
In the following section, the missing mass will be denoted as $M_{0}\left(X^{n},p\right)$
to explicitly show the dependence on the distribution $p$.
\begin{lemma}
\label{lem:Prior1}
For any missing mass estimator $\hat{M}_0(X^n)$ and a random variable
$P$ over a family of distributions $\mathcal{P}$, 
\begin{align*}
&\min_{\hat{M}_{0}}\max_{p\in\mathcal{P}}\mathbb{E}_{X^n\sim p}\left(M_{0}(X^{n},p)-\hat{M}_{0}(X^{n})\right)^{2}\\
&\qquad\qquad\ge \mathbb{E}_{X^{n}\sim P}\left[\mbox{var}_{P|X^{n}}\left[\left.M_{0}\left(X^{n},P\right)\right|X^{n}\right]\right]
\end{align*}
\end{lemma}
\begin{IEEEproof}
\begin{align*}
&\min_{\hat{M}_{0}}\max_{p\in\mathcal{P}}\mathbb{E}\left(M_{0}\left(X^{n},p\right)-\hat{M}_{0}\left(X^{n}\right)\right)^{2}
  \nonumber\\ 
&\geq  \min_{\hat{M}_{0}}\mathbb{E}_{P}\left(\mathbb{E}_{X^{n}|P}\left(\left.M_{0}\left(X^{n},P\right)-\hat{M}_{0}\left(X^{n}\right)\right|P\right)^{2}\right)\\
 & \overset{\left(a\right)}{=}  \min_{\hat{M}_{0}}\mathbb{E}_{X^{n}}\left(\mathbb{E}_{P|X^{n}}\left(\left.M_{0}\left(X^{n},P\right)-\hat{M}_{0}\left(X^{n}\right)\right|X^{n}\right)^{2}\right)\\
 & \overset{\left(b\right)}{=}  \mathbb{E}_{X^{n}\sim P}\left[\mbox{var}_{P|X^{n}}\left[\left.M_{0}\left(X^{n},P\right)\right|X^{n}\right]\right]
\end{align*}
where (a) follows from the law of total expectation and (b) follows
from the fact that (a) is minimized when
$\hat{M}_{0}\left(X^{n}\right)=\mathbb{E}_{P|X^{n}}\left(\left.M_{0}\left(X^{n},P\right)\right|X^{n}\right)$.
\end{IEEEproof}
Lemma \ref{lem:Prior1} gives us a family of bounds depending on the
distribution of the prior $P$. 
The RHS in Lemma \ref{lem:Prior1} can be computed exactly for a
Dirichlet prior with some analysis.
\begin{lemma}
\label{lem:Prior2}
Suppose $P$ has a Dirichlet
distribution $\mbox{Dir}\left(k,\boldsymbol{\alpha}\right)$, where
$\boldsymbol{\alpha}=\left(\alpha_{0},\alpha_{1},\ldots,\alpha_{k-1}\right)$. Then,
we have
\begin{align*}
&\mathbb{E}_{X^{n}}\left[\text{var}_{P|X^{n}}\left[\left.M_{0}\left(X^{n},P\right)\right|X^{n}\right]\right]
  \nonumber\\
&\quad=\frac{B\left(a,n\right)}{\left(a+n\right)^{2}\left(a+n+1\right)}\left(\sum_{u\in\mathcal{X}}\frac{\alpha_{u}\left(a+n\right)-\alpha_{u}^{2}}{B\left(a-\alpha_{u},n\right)}\right.\nonumber\\
&\qquad\left.-\sum_{u\in X}\sum_{v\in\mathcal{X},v\neq u}\frac{\alpha_{u}\alpha_{v}}{B\left(a-\alpha_{u}-\alpha_{v},n\right)}\right),
\end{align*}
where $B\left(\cdot,\cdot\right)$ is the Beta function and $a=\sum_{u\in\mathcal{X}}\alpha_{u}$.
\end{lemma}
We skip the details for want of space.

Let $\boldsymbol{\alpha}=\left(\frac{1}{n},\frac{1}{n},\ldots,\frac{1}{n}\right)$
and $k=cn^{2}$. For this choice of parameters, the expression in
Lemma \ref{lem:Prior2} can be bounded as
\begin{eqnarray*}
\mathbb{E}_{X^{n}}\left[\mbox{var}_{P|X^{n}}\left[\left.M_{0}\left(X^{n},P\right)\right|X^{n}\right]\right] & \geq & \frac{1}{n}\cdot\frac{c}{\left(c+1\right)^{3}}+o\left(\frac{1}{n}\right),
\end{eqnarray*}
where, once again, we skip the details. The coefficient of $\frac{1}{n}$ attains a maximum value of $\frac{4}{27}$
when $c=\frac{1}{2}$, which results in the following bound on the
minimax risk:
\begin{lemma}
\label{lem:Prior3}
\begin{eqnarray*}
\min_{\hat{M}_{0}}\max_{p\in\mathcal{P}}\mathbb{E}\left(M_{0}\left(X^{n},p\right)-\hat{M}_{0}\left(X^{n}\right)\right)^{2} & \geq & \frac{4}{27n}+o\left(\frac{1}{n}\right)
\end{eqnarray*}

\end{lemma}
The bound is worse than the $\frac{1}{4n}$ bound obtained from
distribution estimation in the next section, but it can possibly be
improved by better selection of the prior.

\subsection{Lower bounds via Distribution Estimation}
\label{sec:dist}
To bound the minimax risk for missing mass estimation, one approach
is to reduce the problem to that of estimating a distribution. Let
$\mathcal{P}$ be the set of distributions over the set $\mathcal{X}=\left\{ 0,1\right\} $
such that for all $p\in\mathcal{P}$, $p\left(0\right)\geq\frac{1}{2}$.
A known result (refer \cite{Lehmann1998, Kamath15} for instance) states that
the minimax $\ell^{2}$ loss in estimating $p(0)$ is
$\frac{1}{4n}$. More precisely, let $\hat{p}(X^n)$ be an estimator for
$p(0)$ from a random sample $X^n$ distributed according to $p$. Then,
we have 
\begin{lemma}
\label{lem:DE1}
\begin{eqnarray*}
\min_{\hat{p}\left(0\right)}\max_{p\in\mathcal{P}}\mathbb{E}_{X^n\sim p}\left(p\left(0\right)-\hat{p}\left(X^n\right)\right)^{2} & = & \frac{1}{4n}+o\left(\frac{1}{n}\right)
\end{eqnarray*}
\end{lemma}
For an arbitrary positive integer $k$, let $\mathcal{P}_{c}$ be
the set of distributions over the set $\mathcal{X}=\left\{ 0,1,2,\ldots k-1\right\} $,
such that for any $p_{c}\in\mathcal{P}_{c}$, we have $p_{c}\left(0\right)\geq\frac{1}{2}$
and $p_{c}\left(i\right)=\frac{1-p_{c}\left(0\right)}{k}$ for all
$i\geq1$. We can use Lemma \ref{lem:DE1} to obtain minimax bounds
in estimating $p_{c}\left(0\right)$ for this family of distributions
as well. Let $\hat{p}_c(X^n)$ be an estimator for
$p_c$ from a random sample $X^n$ distributed according to $p_c$. Let
$\hat{p}_c(X^n,i)$ be the probability $\hat{p}_c$ assigns to the symbol $i$.
\begin{lemma}
\label{lem:DE2}
\begin{eqnarray*}
\min_{\hat{p}\left(0\right)}\max_{p\in\mathcal{P}_{c}}\mathbb{E}\left(p_{c}\left(0\right)-\hat{p}_{c}\left(X^n,0\right)\right)^{2} & \geq & \frac{1}{4n}+o\left(\frac{1}{n}\right)
\end{eqnarray*}
\end{lemma}
\begin{IEEEproof}
Suppose we want to estimate an unknown distribution $p\in P$ and
we have an estimator $\hat{p}_{c}$ for distributions in $\mathcal{P}_{c}$.
Then we can use $\hat{p}_{c}$ to estimate $p$ as follows. Take the
observed sample distributed according to $p$, and if it is 0, keep it as it is. If it
is 1, then replace it with an uniformly sampled random variable over
$\left\{ 1,2,\ldots k\right\} $. The result of this sampling process
is a distribution $p_{c}$ in $\mathcal{P}_{c}$ with $p_{c}\left(0\right)=p\left(0\right)$.
Thus, any estimator for distributions in $\mathcal{P}_{c}$ can be
reduced to an estimator for distributions in $\mathcal{P}$ and
\begin{align*}
\min_{\hat{p}\left(0\right)}\max_{p\in\mathcal{P}_{c}}\mathbb{E}_{X^n\sim
  p_c}&\left(p_{c}\left(0\right)-\hat{p}_{c}\left(X^n,0\right)\right)^{2}\\ 
&\geq 
           \min_{\hat{p}\left(0\right)}\max_{p\in\mathcal{P}}\mathbb{E}_{X^n\sim p}\left(p\left(0\right)-\hat{p}\left(X^n\right)\right)^{2}
\end{align*}
and the proof follows from Lemma \ref{lem:DE1}.
\end{IEEEproof}
\begin{lemma}
\label{lem:DE3}Let $k=e^{n}$. With probability at least $1-1/2^n$,
the missing mass $M_{0}\left(X^{n}\right)$ satisfies
$$M_{0}\left(X^{n}\right)  =  1-p\left(0\right)+O\left(ne^{-n}\right).$$
\end{lemma}
\begin{IEEEproof}
Probability of symbol $0$ appearing at least once in $X^n$ is $1-(1-p(0))^n\geq 1-1/2^n$.
 Furthermore,
at most $n$ distinct symbols from $1,2,\ldots k-1$ can appear in
$X^{n}$. Hence, with probability $1-1/2^n$, the observed
mass $1-M_{0}\left(X^{n}\right)$ satisfies
$$p\left(0\right)\leq  1-M_{0}\left(X^{n}\right)  \leq p\left(0\right)+\left(1-p\left(0\right)\right)ne^{-n},$$
and hence follows the lemma.
\end{IEEEproof}
From Lemmas \ref{lem:DE2} and \ref{lem:DE3}, we can obtain a lower
bound of $1/4n$ on the minimax risk of missing mass estimation. Combining the
lower bound with the upper bound on the risk of the Good-Turing
estimator from Theorem \ref{thm:4}, we have the following:
\begin{theorem}
\label{thm:lower_dist}
The minimax risk of missing mass estimation, denoted $R_n^*$, satisfies the following bounds:
\[
\frac{0.25}{n}+o\left(\frac{1}{n}\right) \leq R_n^* \leq \frac{0.6179}{n}+o\left(\frac{1}{n}\right).
\]
\end{theorem}

\section{Summary and Future Directions}
We studied the problem of missing mass estimation and showed that the
minimax risk lies between $0.617/n$ and $1/4n$. We further showed that
the risk of the Good-Turing estimator lies between $0.608/n$ and
$0.617/n$.

Our results pose several interesting questions for future work. Two
natural questions are: (1) are there priors which yield better lower
bounds on the minimax risk of missing mass? and (2) are there
estimators that have better risk than the Good-Turing estimator?

We finally remark that it might be interesting to see if the minimax
risk results imply better concentration results for the missing mass
and the Good-Turing estimator.

\section{Acknowledgements}
Authors thank Alon Orlitsky for helpful discussions.  Ananda Theertha
Suresh thanks Jayadev Acharya for helpful comments.

\bibliographystyle{IEEEtran}
\bibliography{refs}

\end{document}